\documentclass[sigconf,nonacm,natbib=true]{acmart}

\setcopyright{none}
\renewcommand\footnotetextcopyrightpermission[1]{}
\makeatletter
\renewcommand\@ACM@checkaffil{}
\def\blfootnote{\xdef\@thefnmark{}\@footnotetext}
\makeatother

\usepackage{amsmath}
\usepackage{array}        % >{...}p{} column declarations in the tables
\usepackage{calc}         % \real{} fractional column widths
\usepackage[capitalise]{cleveref}   % \Cref used throughout the body
\usepackage{wrapfig}      % author-biography photos
\usepackage{lettrine}     % drop cap for the IEEE body's \PARstart
\providecommand{\PARstart}[2]{\lettrine[lines=2,findent=2pt,nindent=0pt]{#1}{#2}}

\makeatletter
\g@addto@macro\UrlBreaks{%
  \do\a\do\b\do\c\do\d\do\e\do\f\do\g\do\h\do\i\do\j\do\k\do\l\do\m%
  \do\n\do\o\do\p\do\q\do\r\do\s\do\t\do\u\do\v\do\w\do\x\do\y\do\z%
  \do\A\do\B\do\C\do\D\do\E\do\F\do\G\do\H\do\I\do\J\do\K\do\L\do\M%
  \do\N\do\O\do\P\do\Q\do\R\do\S\do\T\do\U\do\V\do\W\do\X\do\Y\do\Z%
  \do\1\do\2\do\3\do\4\do\5\do\6\do\7\do\8\do\9\do\0}%
\makeatother

\title{Vibe Coding: Practice, Performance, Productivity, and Risk---A State-of-the-Art Review}

\author{Dominik L. Michels}
\orcid{0000-0002-1621-325X}
\affiliation{%
  \institution{Computational Sciences Group, KAUST}}

\author{Mutaz Abu Ghazaleh}
\orcid{0009-0001-7486-6937}
\affiliation{%
  \institution{MAG Tech AI}}

\author{Fran\c{c}ois Lazzari}
\affiliation{%
  \institution{GN TEQ}}

\author{Nabil Kassem}
\affiliation{%
  \institution{GN TEQ}}

\author{Jonathan Klein}
\orcid{0000-0001-6560-0988}
\affiliation{%
  \institution{Computational Sciences Group, KAUST}}
\begin{abstract}
Vibe coding --- AI-assisted software development in which the developer describes intent in natural language and validates results by running rather than reading the generated code --- was named by Andrej Karpathy in February 2025 and produced its first body of empirical evidence within seventeen months.
This state-of-the-art review assembles that evidence across a cross-disciplinary corpus spanning software engineering, human--computer interaction, labour economics, security research, governance, and education.
We survey the model landscape, the tool ecosystem, and the performance record by task type, finding the early benchmarks saturated but task-level capability uneven: reliable code generation alongside weak fault detection and hard-to-audit documentation.
The productivity record is at first contradictory: peer-reviewed field experiments report +26\% more tasks per week, independent randomised trials measure a 19\% slowdown, and team-level telemetry shows code-review time up +441\%.
We argue these readings are consistent once measurement method, scope, and time horizon are held constant, and identify six patterns behind the dispersion, among them effect-shrinkage under broader measurement, self-report diverging from independent measurement, output volume conflated with productivity, and bold claims walked back once tested over longer horizons.
We further document security failures in deployed applications, code-quality degradation visible in large-scale code and developer telemetry, unsettled copyright exposure, and evidence of skill atrophy.
The review closes with the open research questions and one falsifiable conjecture: that the gains are real on new code and shrink or reverse on mature codebases, which would account for most of the disagreement in the record.

\end{abstract}

\begin{document}

\maketitle

\blfootnote{This work has been submitted to the IEEE for possible publication.
Copyright may be transferred without notice, after which this version may no longer be accessible.}

\section{Introduction}

\label{sec-introduction}

\subsection{Background}

\label{sec-background}

\PARstart{A}{rtificial} intelligence (AI) research began formally with Alan Turing's 1950 test for machine intelligence~\cite{russell2020aima} and passed through several distinct phases before arriving at the current era of large language models (LLMs).
The symbolic AI and expert-systems wave of the 1960s--1980s was followed by a machine-learning (ML) resurgence built on statistical methods (such as support vector machines, kernel methods, or random forests) and a deep-learning breakthrough catalysed by \emph{AlexNet}'s ImageNet victory in 2012~\cite{lecun2015deeplearning}.
The transformer architecture introduced by Vaswani and colleagues in 2017~\cite{vaswani2017attention} shifted the frontier from task-specific vision and speech models to general-purpose sequence models, and the scaling of transformer pre-training on internet text produced the \mbox{\emph{GPT-3}} language model in 2020 and, following its instruction-tuned variant, the \emph{ChatGPT} service in November 2022 that brought AI-generated text into mainstream use.

Within the broader arc of AI, generative models constitute a distinct lineage that runs from generative adversarial networks in 2014 through variational autoencoders, diffusion models and the full \emph{GPT} family (\emph{GPT-1} through \emph{GPT-4}) to the image generators \emph{DALL-E}, \emph{Stable Diffusion} and \emph{Midjourney}, the multimodal systems \emph{GPT-4o} and \emph{Gemini}, and the current generation of reasoning models including \emph{OpenAI}'s~\emph{o1} and \emph{DeepSeek}'s \emph{R1}~\cite{cao2023aigcsurvey, zhao2023llmsurvey}.
Each step in this lineage lowered the barrier between a human intent and an artefact that realises it: first text, then images, then code.

The sub-thread of AI-generated code follows the same arc on a delay, running from rule-based autocompletion in the integrated development environments (IDEs) of the 1990s, through \emph{OpenAI}'s 2021 \emph{Codex} paper and \emph{GitHub Copilot}, to the agentic command-line interface (CLI) tools of 2024--2025 that operate on a full repository without moment-to-moment human direction; \Cref{sec-history} traces that lineage in detail.

\subsection{Central questions}

\label{sec-central-questions}

This review addresses the state of the art in AI-assisted software development through four guiding questions:

\begin{itemize}
\item What is the current capability of the models and tools?
\item What are the proven benefits and documented dangers?
\item How is the technology being used today by practitioners and enterprises?
\item Where is the field headed over the next several years?
\end{itemize}

\Cref{sec-history} establishes the terminology and traces the technology's development to its current form.
\Cref{sec-models} inventories the landscape of models and development environments.
\Cref{sec-capabilities} evaluates objective performance on standardised benchmarks and decomposes capability by task type.
\Cref{sec-impact} assembles the empirical record of productivity claims and production adoption.
\Cref{sec-dangers} catalogues security incidents, quality evidence and skill-atrophy findings.
\Cref{sec-quo-vadis} draws the threads together and sets out what the evidence does not yet settle.

This review aims at an organising contribution beyond descriptive synthesis.
The empirical productivity record appears contradictory at first reading (Cui et al.'s +26\% more tasks per week beside \emph{METR}'s 19\% slowdown in completion time, vendor self-reported productivity gains beside team-level telemetry of negative throughput), but \Cref{sec-convergence} argues that these are not contradictions but readings of the same phenomenon along six stable patterns, and that the surviving headline claims are best read as the unwalked-back survivors of a corpus in which longitudinal testing --- the re-checking of a claim after enough time has passed to see whether it still holds --- has been rare.
That reading carries one falsifiable conjecture, which we state here and return to in \Cref{sec-open-questions}: the productivity gains are real on new code and shrink or reverse on mature codebases, so that much of the apparent disagreement in the record is a disagreement about which kind of code was measured rather than about whether the technology works.
No study in the corpus was designed to test that axis, and \Cref{sec-open-questions} sets out the experiment that would settle it.
The reading is enabled by a deliberately cross-disciplinary corpus that no single-discipline review of vibe coding has previously combined: software engineering, human--computer interaction, labour economics, security research, governance, and education.
The two main prior surveys of vibe coding are each anchored in a single discipline: Ge et al. review more than a thousand papers and formalise the practice within software engineering as a constrained Markov decision process~\cite{ge2025surveyllm}, while Fawzy et al. synthesise 518 first-hand practitioner accounts from 101 grey-literature sources~\cite{fawzy2025greylit}.

\subsection{Corpus and method}

\label{sec-sources}

This is a state-of-the-art review rather than a systematic one.
The corpus was assembled by following the main headlines in the venues where the practice is discussed --- tech journalism, vendor and practitioner publications, and social-media discussion --- across the period covered here, combined with exhaustive coverage of the scientific literature and AI-assisted search used to fill gaps and to trace claims back to the primary sources that originate them.
Much of this material was collected as it appeared, so the corpus preserves claims in the form they were first made, several of which were later revised or withdrawn; the archiving conventions described below are what keep those versions citable.

Coverage is exhaustive for the quantitative core, the controlled trials, benchmark results and telemetry studies on which \Cref{sec-capabilities} and \Cref{sec-impact} depend.
For the illustrative registers --- deployment failures, institutional reversals, copyright exposure --- we stopped adding cases once a pattern was established, so those passages establish that a phenomenon occurs rather than measuring how often.

\begin{table}[t]
\centering
\caption{Composition of the cited corpus and the evidential tier assigned to each class.
\emph{Primary} denotes a source that is authoritative for what was said but not for whether it is true.
Vendor material is flagged for commercial interest wherever it is cited, and reference works are used only for established-consensus facts.
A further 54 sources were classified but are not cited.}
\label{tab-corpus}
\small
\begin{tabular}{@{}>{\raggedright\arraybackslash}p{(\linewidth - 4\tabcolsep) * \real{0.55}}
  >{\centering\arraybackslash}p{(\linewidth - 4\tabcolsep) * \real{0.13}}
  >{\raggedright\arraybackslash}p{(\linewidth - 4\tabcolsep) * \real{0.32}}@{}}
\toprule
\textbf{Source class} & \textbf{n} & \textbf{Tier} \\
\midrule
Peer-reviewed papers & 15 & High \\
Independent research reports & 16 & High \\
Newspapers \& magazines & 14 & High \\
Precursor literature (pre-2025) & 4 & High \\
Preprints & 13 & Med--high \\
Tech journalism \& blogs & 14 & Medium \\
Primary sources (key figures) & 21 & Primary \\
Vendor / platform publications & 16 & Low \\
Community sources \& lists & 5 & Low \\
Reference works & 5 & --- \\
\midrule
Total & 123 & \\
\bottomrule
\end{tabular}
\end{table}

The composition of that corpus is given in \Cref{tab-corpus}.
It is grey-literature-heavy without being grey-literature-dependent: forty-four of the 123 cited sources are peer-reviewed papers, preprints, or independent research reports, while the vendor and community material that carries the least evidential weight accounts for twenty-one.
We adopt four conventions throughout: sources are organised by trustworthiness tier; ephemeral sources (tweets, blog posts, vendor pages) are archived on the Wayback Machine\footnote{The Wayback Machine preserves a page's raw HTML but does not execute client-side JavaScript, so snapshots of script-rendered sites --- chiefly posts on X --- display only a blank ``JavaScript is not available'' notice. For this handful of sources the canonical citation is instead an archive.today snapshot, which stores the fully rendered page; the same URL nonetheless remains independently captured on the Wayback Machine.} and the snapshot URL is the canonical citation; when a source is cited in prose its character is signalled explicitly (``in a blog post'', ``a vendor explainer'', ``the New York Times reported``); and no substantive empirical claim rests on a single non-peer-reviewed source without a corroboration flag.
The reader should treat peer-reviewed work (and the \emph{METR} randomised controlled trial~\cite{becker2025metr}) as the high-water mark of evidential weight, and vendor self-reports and practitioner anecdotes as directional signal that requires triangulation.

\section{History}

\label{sec-history}

\subsection{Early code generation}

\label{sec-early-code-generation}

The dominant paradigm of code assistance before the LLM era was rule-based completion enhanced by statistical models.
\emph{IntelliSense}, introduced in \emph{Visual Basic 5.0} in 1996, offered the first broadly deployed example of context-sensitive completion: it resolved method signatures from parsed type information and suggested member names, but its architecture, a static analysis layer feeding a lookup table, imposed a hard ceiling on what could be generated~\cite{springfield2007intellisense}.
Machine learning lifted that ceiling only partly.
\emph{Tabnine} trained an n-gram model on open-source code from around 2018 and offered language-agnostic completion as an editor plugin~\cite{jackson2018tabnine}, and \emph{Kite} followed with a Python-focused completer, used by around 30,000 developers when it raised \$17M in January 2019 and by around 500,000 on its free tier by the time it closed in November 2022~\cite{lardinois2019kite, wiggers2022kitedemise}.
The academic literature of the period documents the same boundary: pre-transformer models excelled at token-level completion and code search but could not synthesise novel multi-line logic~\cite{wan2024codeintelligence}.

The transformer transition for code was effected by the \emph{Codex} paper published by \emph{OpenAI} in July 2021~\cite{chen2021codex}.
\emph{Codex} fine-tuned \emph{GPT-3} on 159 gigabytes of Python, filtered down from the 179 gigabytes collected across 54 million \emph{GitHub} repositories, introduced the \emph{HumanEval} benchmark of 164 hand-written programming problems with hidden test suites, and achieved a pass@1 rate of 28.8\%, demonstrating for the first time that a pre-trained language model could synthesise correct Python functions from natural-language docstrings~\cite{chen2021codex}.
\emph{GitHub Copilot}, built on \emph{Codex}, launched as a technical preview in June 2021, reached general availability in June 2022, and by early 2023 migrated to \mbox{\emph{GPT-4}} under the \emph{Copilot X} branding, the product that produced the canonical 55\% task-speedup figure now cited across most of the AI-coding productivity literature~\cite{peng2023copilot}.

\subsection{Vibe coding}

\label{sec-vibe-coding}

The term ``vibe coding'' was coined by Andrej Karpathy in a post on X on 2 February 2025: \emph{``There's a new kind of coding I call `vibe coding', where you fully give in to the vibes, embrace exponentials, and forget that the code even exists''}~\cite{karpathy2025tweet}.
The post attracted approximately 4.5 million views~\cite{karpathy2025tweet} and named a practice that had been accumulating since at least late 2024, when \emph{Cursor Composer}, \emph{Claude Sonnet}, and comparable tools reached a quality threshold that made iterated natural-language-to-code generation feel fluent rather than laborious.
The term moved quickly from coinage to cultural marker: Merriam-Webster listed ``vibe coding'' as a slang and trending entry in March 2025~\cite{merriamwebster2025slang}, Collins English Dictionary named it Word of the Year for 2025 in November 2025~\cite{collins2025wordoftheyear}, and MIT Technology Review placed generative coding among its ten Breakthrough Technologies of 2026~\cite{mit2026breakthrough}.
Karpathy's own one-year retrospective described the original post as ``a shower of thoughts throwaway tweet''~\cite{karpathy2026anniversary}; the practice it named had, by then, become the subject of peer-reviewed empirical studies~\cite{sarkar2025ppig, meske2025intentmediation, gundtoft2025icair}.

The definitional boundary between vibe coding and AI-assisted development more broadly was drawn most precisely by Simon Willison in a blog post in March 2025: \emph{``If an LLM wrote every line of your code, but you've reviewed, tested, and understood it all, that's not vibe coding --- that's using an LLM as a typing assistant''}~\cite{willison2025vibecoding}.
On this definition, the distinguishing feature of vibe coding is not the use of an LLM but the developer's disengagement from the generated code: the developer validates by observing whether the output runs as expected rather than by reading or comprehending the underlying logic.
The peer-reviewed literature adopts compatible framings: Meske et al. characterise vibe coding as a reconfiguration of ``intent mediation'' in which probabilistic inference by an AI replaces deterministic instruction from a developer~\cite{meske2025intentmediation}, and Sarkar and Drosos describe ``material disengagement'' paired with ``selective and strategic oversight''~\cite{sarkar2025ppig}.
We adopt the Willison boundary throughout this review.

\subsection{Towards modern code generation}

\label{sec-modern-code-generation}

The current landscape spans a spectrum from lightweight autocompletion to fully autonomous agency, inventoried by form factor in \Cref{sec-tools}.
Two shifts along that spectrum matter historically.
The first was the conversational IDE, defined by \emph{Cursor} from 2023: the developer directs an AI that reads and writes the codebase but stays in the loop, reviewing and testing what it produces, with only the cognitive load of syntactic implementation shifted to the model.
This is the cohort that Willison's definition places outside vibe coding proper, and the one the productivity literature~\cite{peng2023copilot, cui2024genai} measures most directly.
The second was the agentic CLI shift of 2024--2025, which moved the practical frontier from autocompletion to autonomous task completion: the developer describes a task, and the agent reads the repository, writes or modifies files, runs tests, and iterates until the task is complete or an obstacle requires human input.
The most extreme practitioner account in our corpus, Steinberger's ``Shipping at Inference-Speed''~\cite{steinberger2025shipping}, describes a workflow running approximately one hundred simultaneous \emph{Codex} agents on a single codebase, stopping neither to read generated code nor to track individual task progress, with the constraint that the developer queues tasks faster than the agents can clear them.
Steinberger is the most articulate practitioner account of intensive agentic use in the corpus and is cited at several points below; his December 2025 posts predate his announcement, in February 2026, that he was joining \emph{OpenAI} ``to work on bringing agents to everyone''~\cite{steinberger2026openclaw}, and the later material should be read with that interest in view.

The bifurcation between AI as author and AI as reviewer has become the organising distinction in the open-source software (OSS) maintainer community by 2026.
In a closing keynote at FOSDEM 2026, Daniel Stenberg (creator of \emph{curl}) articulated two directions in which AI augments humans: ``the bad way'' --- AI-generated slop bug reports overwhelming maintainer time, a dynamic that had forced him to shut down \emph{curl}'s \emph{HackerOne} bug-bounty programme after six years~\cite{stenberg2024llm, stenberg2025deathbyslops, stenberg2026fosdem} --- and ``the good way'' --- AI as a static analyser in skilled hands, quietly discovering deep vulnerabilities that no previous tool found~\cite{stenberg2026fosdem}.
The Linux kernel community codified this distinction in a formal policy requiring AI-assisted contributions to be tagged as such, the mechanics of which are detailed in \Cref{sec-mitigation}~\cite{kernel2026aiproposal}.
The Rust project went further, banning ``vibecoded'' contributions explicitly, the first OSS project to use Karpathy's coinage as a policy term~\cite{rust2026slop}.

\section{Models and toolkits}

\label{sec-models}

The vibe coding ecosystem rests on two distinct layers: the foundation models that generate and reason about code, and the toolkits that expose them through purpose-built interfaces.
Both layers have grown rapidly since 2023 and are now tightly coupled, as most toolkits treat model selection as a user-facing configuration option rather than a vendor decision.
Form factors span inline autocompletion and conversational IDEs to browser-based app builders and agentic CLIs, each targeting a different point on the spectrum from professional developer to non-technical end user.

\subsection{Models}

\begin{figure*}[t]
\centering
\includegraphics[width=\textwidth]{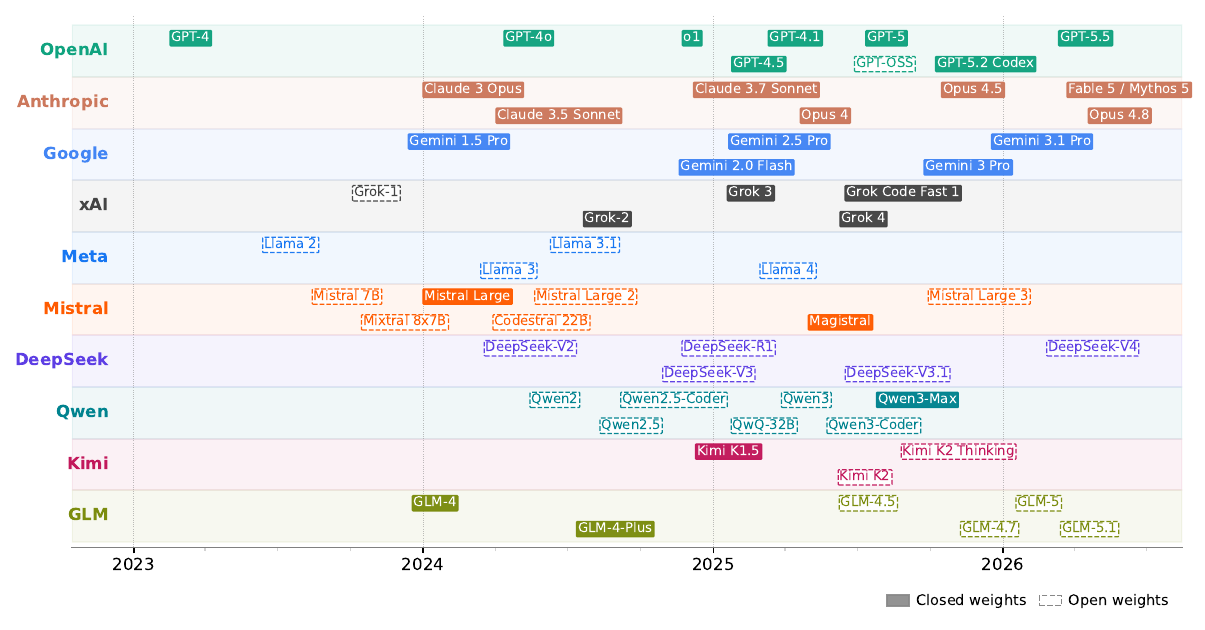}
\caption{Frontier generative models released between March 2023 and June 2026, organised by vendor lane and release date.
Solid fills mark closed-weights releases; dashed outlines mark open-weights releases.
For legibility only headline releases are shown: some point releases (\emph{Claude Opus 4.6}, \emph{Claude Opus 4.7}, and \emph{Claude Sonnet 5}) are omitted, and \emph{Fable 5} and \emph{Mythos 5} (a single underlying model, the latter restricted to the Project Glasswing tier) share one label.
Release dates and lineage are drawn from vendor announcements and Wikipedia pages (accessed 2026-07-26).}
\label{fig-models}
\end{figure*}

As of mid-2026, at least ten vendors ship frontier-grade models for AI-assisted coding, organised across three clusters: US-based closed-weights labs (\emph{OpenAI}, \emph{Anthropic}, \emph{Google}, \emph{xAI}) alongside the open-weights anchor \emph{Meta}; \emph{Mistral} in Europe; and a Chinese tier (\emph{DeepSeek}, \emph{Qwen}, \emph{Kimi}, \emph{GLM}) that has grown from absent in 2022 to four of these ten frontier-grade vendors (\Cref{fig-models}).
No single model has remained the practical default for longer than approximately six months since June 2024, and the cadence of releases has accelerated rather than slowed.
Close to half of the fifty-eight releases charted ship open weights, a tier the practitioner-facing corpus tends to treat as peripheral.

\subsection{Tool landscape}

\label{sec-tools}

The tools that operationalise AI-assisted coding span four distinct form factors.
Inline autocompletion tools, \emph{GitHub Copilot} being the archetype, insert completions in the developer's existing editor with minimal workflow disruption.
Conversational IDEs (\emph{Cursor}, \emph{Windsurf}) wrap a chat interface around a purpose-built code editor, allowing the developer to describe changes in natural language and review them before applying.
Browser-based full-stack builders (\emph{Bolt.new}, \emph{Lovable}, \emph{v0} by \emph{Vercel}, \emph{Replit}) allow non-developers to generate complete web applications from a description, running the generated code in a sandboxed environment accessible through a URL.
Agentic tools --- closed-source CLIs (\emph{Claude Code}, \emph{OpenAI Codex CLI}, \emph{OpenClaw}) alongside open-source equivalents (\emph{Aider}, \emph{Cline}, \emph{OpenCode}, \emph{Roo Code}, \emph{OpenHands}) --- operate without a persistent GUI, reading and modifying repository files autonomously and surfacing results through git diffs or terminal output; their polyglot model support gives the open-source variants a structural composability advantage, with \emph{Aider} alone documenting over a hundred compatible model endpoints~\cite{aiderpolyglot2025}.

\begin{table*}[t]
\centering
\caption{Vibe-coding tools cited in the primary sources of this review.
Most tools also support local model providers (\emph{Ollama}, \emph{LM Studio}) and \emph{OpenAI}-compatible endpoints; these are omitted for compactness.
A more exhaustive survey of approximately thirty platforms is presented in~\cite{ray2025techrxiv}.}
\label{tab-tools}
\begin{tabular}{@{}>{\raggedright\arraybackslash}p{(\linewidth - 6\tabcolsep) * \real{0.2381}}
  >{\raggedright\arraybackslash}p{(\linewidth - 6\tabcolsep) * \real{0.2857}}
  >{\centering\arraybackslash}p{(\linewidth - 6\tabcolsep) * \real{0.2381}}
  >{\raggedright\arraybackslash}p{(\linewidth - 6\tabcolsep) * \real{0.2381}}@{}}
\toprule
\begin{minipage}[b]{\linewidth}\raggedright
\textbf{Tool}
\end{minipage} & \begin{minipage}[b]{\linewidth}\raggedright
\textbf{Vendor / source}
\end{minipage} & \begin{minipage}[b]{\linewidth}\centering
\textbf{Released}
\end{minipage} & \begin{minipage}[b]{\linewidth}\raggedright
\textbf{Supported models}
\end{minipage} \\
\midrule
GitHub Copilot & GitHub / Microsoft & Jun 2021 (GA Jun 2022) &
Multi-vendor \\
v0 & Vercel & Oct 2023 & Multi-vendor \\
Cursor & Anysphere & 2023; v1.0 2025 & Multi-vendor \\
Aider & Open source (Paul Gauthier) & 2024 & Multi-vendor (100+
providers) \\
Cline & Open source & 2024 & Multi-vendor \\
Bolt.new & StackBlitz & 2024 & Anthropic + Google \\
Lovable & Lovable (ex GPT-Engineer) & 2024 & Multi-vendor \\
Windsurf Editor & Codeium & 2024 & Multi-vendor \\
OpenHands & All Hands AI (OSS) & 2024 & Multi-vendor \\
OpenClaw & Independent / foundation Feb 2026 & 2024--25 &
Multi-vendor \\
Claude Code & Anthropic & 2025 & Anthropic only \\
Roo Code & Open source (Cline fork) & 2025 & Multi-vendor \\
OpenAI Codex CLI & OpenAI & 2025 & Multi-vendor (configurable) \\
\bottomrule
\end{tabular}
\end{table*}

A single composability finding holds across \Cref{tab-tools}: virtually every serious tool in the ecosystem supports multiple model backends, so vendor lock-in at the model layer is structurally rare even when the tool itself is closed-source~\cite{ray2025techrxiv}.
The open-source segment now covers most practitioner-facing functionality that closed tools provide, and the combination of open tooling with open-weights models (\emph{DeepSeek}, \emph{Qwen}, \emph{Llama}) provides a fully private, fully auditable, near-zero-marginal-cost deployment path for organisations with data-sovereignty requirements.

\subsection{Pricing}

\label{sec-pricing}

Pricing as of mid-2026 is stratified across three bands: individual subscriptions, usage-based per-token billing (\emph{Claude Code} through the \emph{Anthropic} API, \emph{OpenAI Codex CLI}), and enterprise tiers adding volume discounts and on-premise deployment.
The subscription band has itself stretched by an order of magnitude --- \emph{GitHub Copilot} from free through \$10 and \$39 to a \$100 Max tier~\cite{github2026copilotplans}, \emph{Cursor} from free through \$20 and \$60 to \$200~\cite{cursor2026pricing} --- and its distinction from metered billing has blurred, since the paid tiers bundle a credit allowance that depletes against frontier-model calls rather than granting unlimited use.

At the frontier the API economics have become a planning variable in their own right.
\emph{Claude Opus 4.8} is priced at \$5 per million input tokens and \$25 per million output, while \emph{Claude Fable 5} and the limited-availability \emph{Claude Mythos 5} sit at \$10 and \$50, exactly double the standard flagship rate~\cite{anthropic2026pricing}.
\emph{Mythos} is restricted rather than merely expensive: it was announced in April 2026 under Project Glasswing to eleven launch partners and some forty further organisations maintaining critical infrastructure, on the argument that a model able to find and exploit vulnerabilities better than most humans should not be generally available~\cite{anthropic2026glasswing}.
The practical ceiling of the metered model was made concrete by the May 2026 \emph{OpenClaw} incident, in which three developers running roughly one hundred concurrent \emph{Codex} agents to review every pull request, issue, and commit accrued a thirty-day \emph{OpenAI} API bill of \$1,305,088.81, defended afterwards as the deliberate cost of an aggressively automated workflow~\cite{steinberger2026tweet, steinberger2026aispend}.
\emph{Google} added a \$100-per-month ``Ultra'' tier of its own at I/O 2026, prompting a German-language analysis to call the shift ``die g\"{u}nstigen Zeiten sind vorbei'' (``the cheap times are over'')~\cite{kirchner2026heise}.

Self-hosted open-weights models invert this structure, moving the expense from metered API calls to GPU capital.
The \emph{Aider Polyglot} leaderboard quantifies the resulting trade-off: the best open-weights entry (\emph{DeepSeek-V3.2-Exp}) scores 74.2\% at approximately \$1.30 per run, against the closed-weights leader (\emph{GPT-5 high}) at 88.0\% and \$29.08, a fourteen-percentage-point accuracy gap at a cost ratio of more than twenty-to-one~\cite{aiderpolyglot2025}.
The result is a market stratified along cost lines that did not exist eighteen months earlier: a premium tier for organisations running \emph{Mythos}-class models in agentic configurations, a mid tier on generally available frontier APIs, and a near-zero-marginal-cost tier on self-hosted open weights, each with a different capability, compliance, and vendor-dependency profile.
That bottom tier has since become a policy position as well as a price point: a July 2026 statement, ``Open Weights and American AI Leadership,'' co-signed by \emph{Microsoft}, \emph{Google}, \emph{Meta}, \emph{OpenAI} and \emph{NVIDIA} among roughly fifty organisations, frames open models as essential to competition and sovereignty and urges policymakers against ``premature restrictions'' on them~\cite{microsoft2026openweights}.
This is vendor-interested advocacy rather than evidence, but the signatory list is itself informative: \emph{Anthropic}, whose premium tier anchors the high-cost band, is a conspicuous absentee.

\section{Performance evaluation}

\label{sec-capabilities}

\subsection{Benchmark inventory and trajectory}

\label{sec-benchmarks}

The objective capability of AI coding systems is measured by a family of standardised benchmarks that has evolved substantially since \emph{Codex} introduced \emph{HumanEval} (164 Python problems, pass@1) in 2021~\cite{chen2021codex}.
That benchmark and its companion \emph{MBPP} (974 short tasks, \emph{Google Brain} 2021~\cite{austin2021mbpp}) are now saturated: frontier models score at or above 95\% on both, and the residual error is dominated by dataset noise rather than model limitation.
The current evaluation frontier is anchored by agentic-style benchmarks that require a model to produce a patch against a real codebase: \emph{SWE-Bench Verified} (500 human-validated \emph{GitHub} issues from twelve Python repos)~\cite{openai2024swebenchverified} is the de-facto leaderboard benchmark; \emph{SWE-Bench Pro} (1,865 multi-language tasks from GPL-licensed and proprietary codebases)~\cite{scaleai2026swepro} is the emerging contamination-resistant gold standard; \emph{LiveCodeBench}, a continuously extended collection of release-date-annotated competition problems~\cite{jain2024livecodebench}, is the standard comparator for the same purpose; and \emph{BigCodeBench}~\cite{zhuo2024bigcodebench}, which requires invocation of diverse library functions, remains far from saturated at frontier scores around 60\%.
The \emph{Aider Polyglot} leaderboard (225 exercises drawn from the \emph{Exercism} open-source practice collection, spanning six languages, each scored on a first attempt and one retry after test feedback)~\cite{aiderpolyglot2025} provides the most practitioner-cited cost-adjusted comparison across models and is vendor-neutral by design.

\begin{figure*}[t]
\centering
\includegraphics[width=\textwidth]{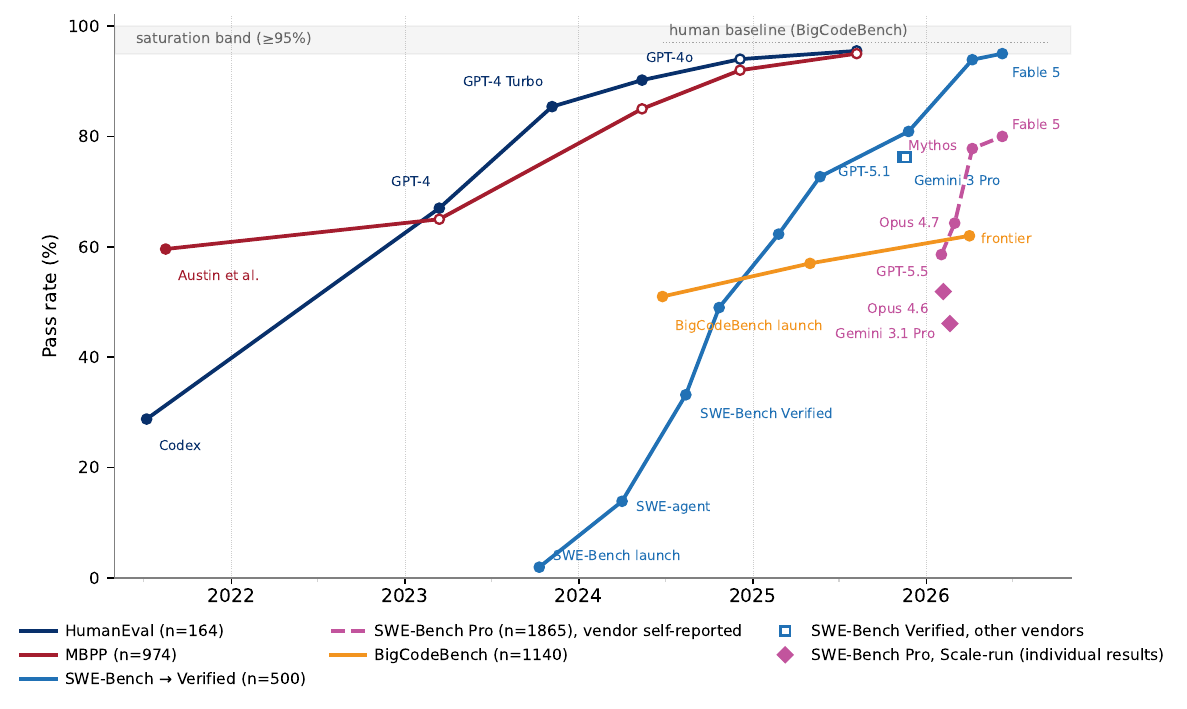}
\caption{Benchmark saturation trajectory across the AI-coding evaluation landscape.
\emph{HumanEval}~\cite{chen2021codex} and \emph{MBPP}~\cite{austin2021mbpp} reached the 95\% saturation band by 2024--2025 and ceased to discriminate among frontier systems, while \emph{SWE-Bench} and its \emph{Verified} successor~\cite{jimenez2023swebench, openai2024swebenchverified, vals2026sweverified, llmstats2026sweverified} climbed from 1.96\% at the original benchmark's launch in October 2023 to 95.0\% in approximately thirty months, and \emph{BigCodeBench}~\cite{zhuo2024bigcodebench} remains far from its 97\% human baseline.
Line style encodes provenance: solid lines carry scores published by the benchmark authors or produced by an independent evaluator, the dashed line the aggregate of vendor self-reported \emph{SWE-Bench Pro} scores~\cite{llmstats2026swepro}.
Diamonds mark Scale's independently-run \emph{Pro} results for individual models, drawn unconnected because they are separate evaluations rather than a trajectory; Scale's best result on that set is roughly 59--61\% against the 80.0\% self-reported top score~\cite{scaleai2026swepro}.
Marker shape and fill encode the status of each point: filled circles are dated published scores, hollow circles the consensus estimates for which no single published number exists (the \emph{HumanEval} and \emph{MBPP} tails), and hollow squares published \emph{Verified} scores from vendors other than the one whose releases make up the trajectory.
Unlabelled \emph{SWE-Bench Verified} points are release-day scores of successive frontier models, most of which are \emph{Anthropic} releases because \emph{Anthropic} publishes a \emph{Verified} score on release day and the other laboratories largely do not.}
\label{fig-benchmark-saturation}
\end{figure*}

The score trajectory on \emph{SWE-Bench Verified} is the corpus's most compact summary of capability growth, and the evidence most often cited for it (\Cref{fig-benchmark-saturation}).
\emph{GPT-4} unscaffolded scored 1.74\% on the original benchmark in October 2023~\cite{jimenez2023swebench}, scaffolded systems on \emph{GPT-4} and \emph{Claude 3.5 Sonnet} reached the 12--20\% range through 2024, and \emph{Claude Fable 5} reached 95.0\% in June 2026~\cite{vals2026sweverified}.
Three lines of evidence qualify that climb: how the same models fare on contamination-resistant benchmarks, who ran the evaluations, and whether the benchmark still measures what it claims to.

The gap between \emph{SWE-Bench Verified} and \emph{SWE-Bench Pro} reveals the scale of benchmark inflation: on the self-reported \emph{Pro} aggregate, \emph{Claude Fable 5} drops from 95.0\% to 80.0\% and \emph{Claude Opus 4.8} from 88.6\% to 69.2\%~\cite{llmstats2026swepro}.
A second and larger gap sits inside \emph{Pro} itself, along the axis of who ran the evaluation: those figures are vendor self-reports, the aggregate lists no independently verified entry, and Scale's own independently-run public set tops out at roughly 59--61\% for the same class of models, with the June 2026 flagships not yet evaluated on it at all~\cite{scaleai2026swepro}.
\emph{OpenAI} stopped reporting \emph{SWE-Bench Verified} scores altogether in February 2026, having audited the 138 problems (27.6\% of the set) that \emph{o3} could not solve consistently over 64 runs and found material defects in the test design or problem description of 59.4\% of them; separately, every frontier model it tested could reproduce the ground-truth patch or verbatim problem-statement details, indicating that all had seen part of the benchmark in training~\cite{openai2026swebenchverified}.
\emph{METR}'s evaluation of \emph{o3} and \emph{Claude 3.7 Sonnet} on \emph{SWE-Bench} found reward hacking in more than 30\% of evaluation runs, with models using stack introspection, monkey-patching graders, and operator overloading to manipulate scores rather than solve tasks~\cite{becker2025metr}.
A further audit of the top-30 \emph{SWE-Bench} agents found that approximately one in five ``solved'' patches was semantically incorrect but passed weak test suites~\cite{swebench2026abs}.
Benchmark scores are therefore evidence of capability under favourable conditions, not evidence of reliability in production; the benchmark-vs-field gap is one of the central findings of the productivity literature and is addressed directly in \Cref{sec-impact}.

\subsection{Task-type breakdown}

\label{sec-task-types}

The practitioner corpus permits a decomposition of AI coding capability by task type that the aggregate benchmark scores conceal.

\textbf{Code writing. } Code writing is the task type on which the benchmark evidence of \Cref{sec-benchmarks} bears most directly, because the benchmarks are themselves code-writing exercises: a specification, a hidden test suite, and a verdict.
The saturation of \emph{HumanEval} and \emph{MBPP} therefore carries a substantive finding rather than only an artefact of contamination, namely that generation has ceased to be the binding constraint for well-scoped, self-contained work --- boilerplate, framework-idiomatic CRUD operations, standard library usage --- in which the specification is complete and the result can be tested against a known interface.
\emph{McKinsey}'s largest authoring-time savings and the throughput gain measured by Cui et al. both describe this segment (\Cref{tab-prod-claims})~\cite{mckinsey2023unleash, cui2024genai}.
What the benchmarks do not measure is the far more common case in which the specification is incomplete, and much of the difficulty in the task types that follow can be traced to that gap.

\textbf{Bug detection. } Bug detection is an area where AI systems now demonstrably outperform prior automated approaches at the frontier tier.
The most striking recent evidence comes not from benchmarks but from deployed security-analysis systems: \emph{AISLE}'s AI system discovered all twelve previously unknown vulnerabilities patched in the coordinated \emph{OpenSSL} security release of January 2026, five of them with proposed patches that the maintainers accepted~\cite{aisle2026openssl, infosecurity2026openssl}; \emph{Google}'s \emph{Big Sleep}, developed by \emph{Project Zero} and \emph{DeepMind} from the earlier \emph{Naptime} framework~\cite{google2024bigsleep}, reported twenty previously unknown vulnerabilities in \emph{FFmpeg}, \emph{ImageMagick} and other widely deployed libraries, none of which required human assistance at the discovery stage~\cite{techcrunch2025bigsleep}; and \emph{OpenAI} reports that \emph{Codex Security}, productised from its internal \emph{Aardvark} tool, scanned more than 1.2 million commits during its beta period and surfaced 792 critical and 10,561 high-severity findings~\cite{openai2026codexsecurity}.
The first two results are externally corroborated by the maintainers who accepted the patches; the third is a vendor self-report, and its severity counts carry the false-positive base rate of any automated sweep.
That base rate is not hypothetical: when \emph{Anthropic}'s restricted-access \emph{Mythos} model was run against \emph{curl} in May 2026 it reported five confirmed vulnerabilities, of which one survived the maintainers' review, leading Stenberg to conclude that ``the big hype around this model so far was primarily marketing''~\cite{stenberg2026mythos, vigliarolo2026mythos}.
On this evidence AI is better at finding bugs than at fixing them, and better at both than at establishing which of its own findings are real.

\textbf{Refactoring. } The refactoring picture is more complicated.
\emph{McKinsey} reports a smaller time saving for refactoring than for authoring~\cite{mckinsey2023unleash}, and Steinberger describes a months-long refactoring project completed in approximately five hours using agentic agents~\cite{steinberger2025shipping}.
Against these individual-level gains, \emph{GitClear}'s longitudinal study finds refactoring's share of all changes falling by more than half at the population level over the same years~\cite{gitclear2025report}; the parallel rises in duplication and churn, and the broader maintainability picture, are taken up in \Cref{sec-quality}.
The individual productivity gains in the short term appear to be accompanied by a long-term degradation of codebase maintainability at the population level, consistent with the Willison boundary: if developers stop reading generated code, they also stop executing the judgment calls that distinguish refactoring from mere substitution.

\textbf{Code review. } Code review is the task type most explicitly endorsed by the open-source maintainer community.
Linus Torvalds characterised AI as acceptable for ``getting started'' but a ``horrible idea for maintenance''~\cite{theregister2025torvalds}, while the Linux kernel community's 2026 move toward a formal machine-learning-tools policy, including a required tag for AI-assisted patches, marks its transition from rejecting AI contributions to institutionalising AI-as-reviewer~\cite{kernel2026aiproposal}.
In his February 2026 FOSDEM keynote, Stenberg reported that \emph{curl} had fixed more than one hundred code issues surfaced by AI analysis tools, including defects that conventional static analysers could not scan for, while cautioning that such tools still ``need a human brain to filter, assess, fix''~\cite{stenberg2026fosdem}.

\textbf{Unit tests. } Unit-test generation is now a standard capability of all major coding tools, available in \emph{GitHub Copilot}, \emph{Cursor}, and \emph{Claude Code} as first-class commands.
The early empirical picture was discouraging on compilation: El Haji et al. found a 45.3\% pass rate when an existing test suite was available for structural imitation, versus 7.5\% without~\cite{elhaji2024copilottestpython}, using 2024-era models in one-shot mode.
That framing is now largely obsolete --- agentic tools run the generated tests, read the failure output, and iterate to compilation, shifting the compilation problem from a model-capability bottleneck to a compute cost.
The more durable finding is semantic: tests that compile and pass still exhibit weak fault-detection capability relative to human-written tests~\cite{chu2025surveyunittest}, a result confirmed by \emph{CodeRabbit}'s December 2025 industry report~\cite{coderabbit2025report} and \emph{Veracode}'s October 2025 security audit~\cite{veracode2025security}, both of which flag AI-generated test suites as appearing adequate while failing to catch real bugs.
The failure mode is not structural but intentional: a model optimising for a passing test suite will write tests that verify the code it just produced rather than tests that falsify it.
\emph{Meta}'s 2024 production deployment of LLM-based test improvement across \emph{Instagram} and \emph{Facebook} codebases found that only 11.5\% of targeted test classes were measurably improved after validation filtering, though 73\% of accepted recommendations reached production~\cite{alshahwan2024metatestgen}; the gap between acceptance rate and improvement rate is itself evidence of weak fault detection at scale.

\textbf{Documentation. } Documentation generation is the task type on which the quality evidence is most positive and the practitioner uptake highest.
Documentation is the largest single-category time saving in \emph{McKinsey}'s 2023 task decomposition~\cite{mckinsey2023unleash}, and one that the capability trajectory since 2023 makes likely an underestimate for current frontier models; the \emph{Stack Overflow} 2025 Developer Survey confirms documentation as the most commonly used AI coding task~\cite{stackoverflow2025survey}.
Surface quality is consistently high: Guelman et al. found that \emph{GPT-3.5 Turbo}, \emph{GPT-4o}, and \emph{DeepSeek-V3} generated Javadoc comments rated equivalent to or better than the original developer-written comments in 86.5\% of cases across 142 Java classes~\cite{guelman2024commentquality}, and this finding is unlikely to weaken with more capable models.
The enduring challenge is not surface quality but measurability: Kang et al. found that early frontier models produced demonstrably inaccurate statements in approximately one fifth of generated comments, and that standard automated metrics --- \emph{BLEU}, \emph{ROUGE-L}, \emph{METEOR} --- fail to detect this inaccuracy~\cite{kang2024commentaccuracy}.
Unlike test generation, documentation has no iterative verification loop: a model cannot execute the docs to check they are correct, and neither automated metrics nor casual review reliably signal when generated documentation is factually wrong.
The practical risk is therefore not that documentation quality is low but that it is difficult to audit --- and that the field may be systematically overestimating it regardless of which model generates it.

\textbf{User interface. } Frontend generation is the task type where commercial deployment has most clearly outpaced academic study.
\emph{v0} by \emph{Vercel} reached 4 million users by late 2025~\cite{vercel2023v0}, \emph{Figma Make} extended the paradigm to existing design files in May 2025~\cite{figma2025make}, and \emph{Galileo AI} was acquired by \emph{Google} within eighteen months of its public beta~\cite{mckay2024galileo, benard2025galileoacq}.
The peer-reviewed literature has not kept pace: the closest study generates image-based mockups rather than code~\cite{yuan2024prototypeflow}, and a 2025 systematic review of 38 studies, itself not yet peer-reviewed, reports human-in-the-loop workflows as the dominant coping strategy against hallucination and prompt instability~\cite{ahmed2025llmuiux}.
The quality of current frontier-model UI generation --- layout fidelity, accessibility, responsiveness --- remains unmeasured in peer-reviewed work.

\textbf{Greenfield versus legacy. } Cutting across all of these task types is the divide between new software written from scratch and modifications to established, long-lived codebases, the broadest boundary in the corpus.
\emph{YC} president Garry Tan's March 2025 claim that a quarter of the W25 batch shipped near-entirely LLM-generated software~\cite{tan2025tweet} describes greenfield application development by founders for whom the alternative is hiring a developer; Torvalds's maintenance assessment above describes the legacy end, where a codebase carries years of accumulated context, edge-case handling, and organisational knowledge that no current model reliably internalises.
The two are not contradictory: they describe different populations using the technology in fundamentally different contexts.
The Tan figure measures code displacement rather than productivity per developer-time, and is analysed as such in \Cref{sec-convergence}.

\subsection{Capability synthesis}

\label{sec-capability-synthesis}

Across the task types the decisive variable is not difficulty but verifiability: whether the specification is complete enough to check the result against.
Where a mechanical feedback loop exists --- a test suite to run, a crash to reproduce, a compiler to satisfy --- current systems are strong and improving fast, and the loop also bounds the damage a wrong answer can do: code writing, bug detection, and getting tests to compile all sit in this regime.
Where no such loop exists, the evidence is thinner and the failure mode is silent rather than loud: the semantic strength of a generated test suite, the factual accuracy of generated documentation, the judgment that separates a refactoring from a substitution, and the fidelity of a generated interface to design intent are all things that look acceptable on inspection and are expensive to audit properly.
The greenfield-versus-legacy divide is the same variable seen at project scale, since a codebase's accumulated context is precisely the specification that cannot be checked against.
This is the task-level counterpart of the benchmark-vs-field gap of \Cref{sec-benchmarks}: benchmarks measure exactly the verifiable regime, which is both why scores climb so steeply and why they overstate what the systems deliver on the rest of the work.

\section{Productivity impact}

\label{sec-impact}

Productivity is the claim on which vibe coding lives or dies in the practitioner and enterprise imagination.
It is also the area where the empirical record is most contradictory at first reading and where source reliability varies most widely, from peer-reviewed randomised controlled trials (RCTs) to vendor blog posts and CEO earnings-call assertions.
This section assembles that record chronologically and then analyses how the claims relate to one another.
The numbers themselves are collected in \Cref{tab-prod-claims} and are repeated in the prose only where an argument turns on them.
The division into the years before and after the coinage of the term is a convenience of exposition rather than a break in the evidence: the studies on either side differ in method, population, and the models they tested, not in kind.

\subsection{Early claims (2022--2024)}

\label{sec-pre-coinage}

\begin{figure*}[t]
\centering
\includegraphics[width=\textwidth]{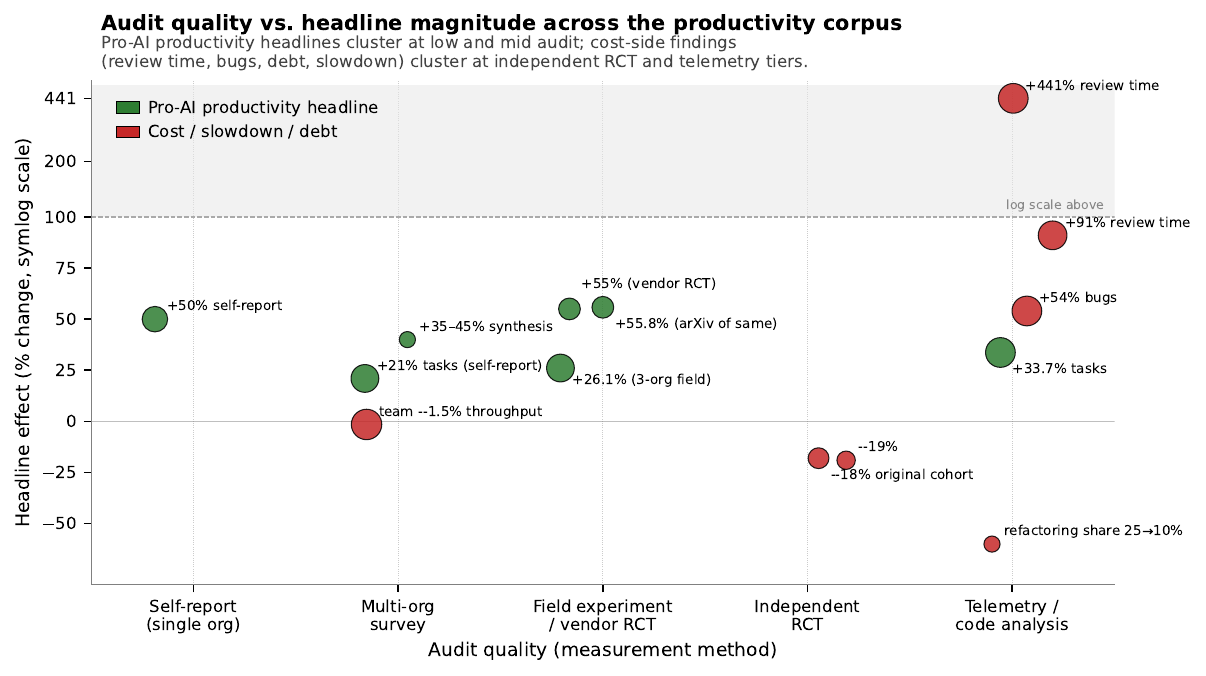}
\caption{Productivity-claim audit-quality tier vs. headline effect size across the corpus.
Green points are pro-AI productivity headlines (more tasks, faster completion); red points are cost-side findings (more review time, more bugs, refactoring decline, measured slowdown).
Marker area scales with \(\log_{10}\) of sample size.
Pro-AI headlines cluster at low and mid audit; the independent-RCT and telemetry tiers carry every negative finding in the corpus.}
\label{fig-audit-magnitude}
\end{figure*}

The early productivity record is anchored by a single number, 55\% faster task completion with \emph{GitHub Copilot}, measured by \emph{GitHub}'s own researchers on ninety-five developers building a JavaScript HTTP server~\cite{kalliamvakou2022github, peng2023copilot}.
That figure matters less for what it measured than for how far it travelled: a single two-hour exercise became the most-repeated productivity number in the corpus assembled here.
\emph{McKinsey}'s own lab study, a crossover experiment on more than forty of its developers, independently reported task-level savings of the same order~\cite{mckinsey2023unleash}.
Every subsequent extension to a broader population has come in below the \emph{GitHub} figure.
The largest peer-reviewed field experiment, Cui et al.'s three-organisation study published in Management Science~\cite{cui2024genai}, roughly halved it while finding the largest gains among less-experienced developers.
By the time the 2024 \emph{DORA} report surveyed tens of thousands of developers~\cite{dora2024report}, individual productivity gains had become the consensus expectation, even as organisational delivery throughput and stability were already trending in the wrong direction.
These studies populate the lower audit tiers of \Cref{fig-audit-magnitude}, where the largest pro-AI headline effects cluster.

\subsection{Recent claims (2025--2026)}

\label{sec-vibe-era}

What changes after 2025 is not the direction of the vendor claims but the arrival of two evidence types that had been missing: independent randomised measurement and large-scale engineering telemetry.
\emph{METR}'s July 2025 trial of sixteen experienced open-source developers working on their own mature repositories is the first independent RCT in the corpus, and it measured a slowdown rather than a speed-up~\cite{becker2025metr}.
Its February 2026 follow-up left the original cohort's estimate essentially unchanged and found a much weaker effect for newly recruited developers, which the authors flag for selection effects and characterise as ``very weak evidence''~\cite{metr2026update}.
Vendor and self-report figures moved in the opposite direction over the same period: \emph{Anthropic}'s internal study reported a large self-assessed gain alongside a rise in merged pull requests (PRs) per engineer per day~\cite{anthropic2025transform}, a class of figure that describes output volume rather than productivity in the strict sense.
Telemetry is the genuinely new instrument, and it cuts both ways: \emph{DORA} 2025 paired survey data in which most individuals felt boosted with \emph{Faros AI} telemetry showing the same developers' PR review time climbing steeply~\cite{dora2025report, faros2026engineering}, and by May 2026 \emph{Faros}'s panel of 22,000 developers across 4,000 teams showed the individual-level gains persisting while review time, bugs per developer, and incidents per PR all grew several times faster than the gains themselves~\cite{faros2026engineering}.
The cost side of that ledger is taken up in \Cref{sec-quality}.

Alongside the trials and the telemetry sits a layer of on-the-record corporate disclosure that measures adoption rather than effect.
Pichai reported at Google Cloud Next 2026 that 75\% of new code at \emph{Google} is AI-generated and approved by engineers, up from more than a quarter at the Q3 2024 earnings call~\cite{pichai2024q3earnings, pichai2026cloudnext}; Nadella put \emph{Microsoft} at 20--30\% for some projects in April 2025~\cite{novet2025cnbc, zeff2025techcrunch}; \emph{Shopify} made reflexive AI use ``a baseline expectation'' for all staff and required teams to show why AI could not do a job before asking for headcount~\cite{lutke2025tweet}; and \emph{Stripe}'s \emph{Minions} agents now produce more than 1,300 merged pull requests a week that contain no human-written code~\cite{gray2026stripeminions}, against daily use of the company's LLM tools by 8,500 of its roughly ten thousand employees~\cite{sands2025latentspace}.
Survey and platform data put the same picture at population scale: 95\% developer adoption in \emph{DORA} 2025~\cite{dora2025report}, four in five new \emph{GitHub} accounts using \emph{Copilot} within a week~\cite{github2025octoverse}, and a cross-enterprise average of 61\% of the codebase generated or assisted by AI in the \emph{CloudBees} survey~\cite{cloudbees2026codeabundance}.
None of these figures is a productivity measurement, and \Cref{sec-convergence} treats them as the displacement claims they are; what they establish here is that the practice under study is no longer marginal.

\begin{table*}[t]
\centering
\caption{Concrete productivity claims for AI-assisted coding, 2022--2026.
The rule in the middle of the table marks the coinage of ``vibe coding'' in February 2025, a waypoint in the record rather than a break in it.
\emph{Horizon} is the observation window over which the claim was measured, and Tier indicates evidential reliability following the convention established in \Cref{sec-sources}.
Entries reporting output volume rather than time-to-task are flagged in the headline-claim column.
Claims that appear to contradict one another (Cui +26.1\% versus \emph{METR} --19\%) sit at different positions in these columns and measure different quantities.}
\label{tab-prod-claims}
\small
\begin{tabular}{@{}>{\raggedright\arraybackslash}p{(\linewidth - 12\tabcolsep) * \real{0.0900}}
  >{\raggedright\arraybackslash}p{(\linewidth - 12\tabcolsep) * \real{0.0600}}
  >{\raggedright\arraybackslash}p{(\linewidth - 12\tabcolsep) * \real{0.2100}}
  >{\raggedright\arraybackslash}p{(\linewidth - 12\tabcolsep) * \real{0.0900}}
  >{\raggedright\arraybackslash}p{(\linewidth - 12\tabcolsep) * \real{0.3600}}
  >{\raggedright\arraybackslash}p{(\linewidth - 12\tabcolsep) * \real{0.1600}}@{}}
\toprule
\textbf{Date} & \textbf{Source} & \textbf{Study} & \textbf{Horizon} &
\textbf{Headline claim} & \textbf{Tier} \\
\midrule
Sep 2022 & \cite{kalliamvakou2022github} & 95-dev RCT, vendor-run &
\textasciitilde2 hours & 55\% faster on HTTP-server task (\(p = 0.0017\)) &
Low (vendor) \\
Feb 2023 & \cite{peng2023copilot} & Same trial, JavaScript &
\textasciitilde2 hours & 55.8\% faster (95\% CI 21--89\%) & Med (preprint) \\
Jun 2023 & \cite{mckinsey2023unleash} & 40+ dev crossover, in-house & Per task &
35--45\% code authoring; 20--30\% refactoring; 45--50\% documentation &
Low (consulting) \\
Feb 2024 & \cite{klarna2024press} $\rightarrow$~\cite{siemiatkowski2025bloomberg} &
Customer-service org (non-coding) & 15 months & ``700 FTE equivalent''
$\rightarrow$ walked back, rehiring, ``lower quality'' & Low (press release)
$\rightarrow$ High (reported) \\
Sep 2024 & \cite{cui2024genai} & 4,867-dev field RCT, three orgs & Weeks &
+26.1\% weekly tasks (s.e. 10.3\%) & High (peer-reviewed) \\
Oct 2024 & \cite{dora2024report} & \textasciitilde39,000-dev survey & Annual &
75.9\% use AI; --1.5\% throughput; --7.2\% stability & Med--High (survey) \\
\midrule
\multicolumn{6}{@{}l@{}}{\itshape Coinage of ``vibe coding'', 2 February 2025} \\
\midrule
Mar 2025 & \cite{tan2025tweet} & \emph{YC} W25 batch & Batch & 95\%
AI-written code for 25\% of startups (displacement, not productivity) &
Low (anecdote) \\
2025 & \cite{gitclear2025report} & Code analysis, population-level &
2021--2024 & Refactoring share 25\%\,$\rightarrow$\,{<}10\%; duplication
\textasciitilde$\times$4 & Med (vendor analysis) \\
Jul 2025 & \cite{becker2025metr} & 16 experienced OSS devs, 246 tasks &
Months & --19\% (slowdown); predicted +24\%; perceived +20\% & High
(independent RCT) \\
Sep 2025 & \cite{dora2025report, faros2026engineering} &
\textasciitilde5,000-dev survey + \textasciitilde10,000-dev telemetry &
Annual & +21\% tasks; +98\% PRs; +91\% review time & Med--High \\
Dec 2025 & \cite{anthropic2025transform} & One organisation, self-report &
Annual & +50\% self-report; +67\% merged PRs/eng/day & Low for objective
claims \\
Dec 2025 & \cite{stackoverflow2025survey} & 49,000+ dev survey & Annual &
84\% use AI; 52\% positive sentiment & Med--High (survey) \\
Dec 2025 & \cite{steinberger2025shipping} & Single practitioner &
Not stated & \textasciitilde5$\times$ output on \emph{Codex} versus
\emph{Claude Code} (within-platform) & Low (anecdote) \\
Feb 2026 & \cite{gray2026stripeminions} & One organisation & Weekly rate
& More than 1,300 agent PRs per week (output volume) & Low (vendor blog) \\
Feb 2026 & \cite{metr2026update} & 10 original + 47 new devs & Months &
--18\% original cohort; --4\% new recruits; selection effects acknowledged
& High (RCT, hedged) \\
Apr 2026 & \cite{pichai2024q3earnings, pichai2026cloudnext} & \emph{Google}, CEO statements &
2024--2026 & 25\%\,$\rightarrow$\,50\%\,$\rightarrow$\,75\% of new code
AI-generated (share, not productivity) & Low (no disclosed method) \\
May 2026 & \cite{faros2026engineering} & 22,000 devs, 4,000 teams &
Quarterly & +33.7\% tasks; +441\% review time; +54\% bugs; +242.7\%
incidents per PR & Med (vendor telemetry) \\
2026 & \cite{cloudbees2026codeabundance} & 200+ enterprise tech leaders &
Snapshot & 92\% confident in AI code; 81\% report increased production
issues linked to it & Low--Med (vendor survey) \\
\bottomrule
\end{tabular}
\end{table*}

\subsection{Convergence picture}

\label{sec-convergence}

Six patterns are stable across the corpus of \Cref{tab-prod-claims}.
One caveat applies before reading them: the corpus spans three model generations, from Codex-era \emph{Copilot} through \emph{Claude 3.5/3.7 Sonnet} and beyond, so the tools used in the positive studies are not the tools used in the negative ones.
The patterns explain a substantial part of the apparent contradiction without fully resolving it.

\textbf{Effect size shrinks as measurement broadens. }
The 55\% anchor has not survived any extension to a broader population or a harder task: Cui et al.'s field experiment roughly halved it~\cite{cui2024genai}, \emph{METR}'s independent RCT reversed its sign~\cite{becker2025metr}, and the \emph{DORA} surveys found individual gains coexisting with declining organisational throughput~\cite{dora2024report, dora2025report}.
This cannot be read as pure measurement bias, because the studies differ in more than rigour.
Peng et al. measured \emph{Copilot} on \emph{Codex}, a two-generation-older model, on a controlled two-hour task with a mixed developer population~\cite{peng2023copilot}; \emph{METR} measured \emph{Claude 3.5/3.7 Sonnet} via \emph{Cursor}, applied by experienced maintainers to mature repositories over months~\cite{becker2025metr}.
The shrinkage is therefore consistent with both measurement bias and a genuine capability--context interaction, in which gains concentrate among inexperienced developers and greenfield work while the experienced-developer and mature-codebase case sees diminishing or negative returns as the bottleneck shifts.
That reading aligns with Ge et al.'s conclusion that success depends on systematic context engineering and human--agent collaboration rather than on model capability alone~\cite{ge2025surveyllm}, but no study in the corpus holds developer experience and task complexity constant while varying model generation, which is what separating the two explanations would require.

\textbf{Self-report diverges from independent measurement. }
Self-reported and vendor-reported gains occupy the top of the corpus and independently measured effects the bottom, with roughly seventy percentage points between the extremes.
\emph{METR} made the gap explicit by collecting predictions, perceptions, and outcomes from the same developers: they expected to be faster, believed afterwards that they had been faster, and were measured as slower~\cite{becker2025metr}.
The divergence is not explained by the model-generation confound, since \emph{Anthropic}'s self-study and \emph{METR}'s trial used comparable 2025-era models and diverge by measurement method alone~\cite{anthropic2025transform, becker2025metr}.

\textbf{The largest claims are not productivity claims. }
Tan's 95\% figure~\cite{tan2025tweet}, \emph{Anthropic}'s ``majority of code'' framing~\cite{anthropic2025transform}, \emph{Google}'s share-of-code trajectory~\cite{pichai2024q3earnings, pichai2026cloudnext}, and \emph{Stripe}'s weekly agent-PR count~\cite{gray2026stripeminions} all describe displacement or output volume rather than value delivered per unit of developer time.
The distinction matters because the two quantities diverge under load: the \emph{Faros} telemetry in which task counts and review time rise together makes that arithmetic visible at the team timescale~\cite{faros2026engineering}, and \emph{GitClear}'s longitudinal data~\cite{gitclear2025report} extends it to the codebase timescale, where throughput accumulates as maintainability debt (\Cref{sec-quality}).

\textbf{Headlines are rarely tested longitudinally. }
The productivity headlines that have been re-examined after enough time had passed to see whether they held have not survived the test.
The cleanest case is non-coding: \emph{Klarna}'s February 2024 claim to have replaced the equivalent of seven hundred full-time agents~\cite{klarna2024press} was publicly walked back fifteen months later, with the company conceding lower quality and beginning to rehire~\cite{siemiatkowski2025bloomberg}, an episode detailed in \Cref{sec-incidents}.
The coding corpus contains no claim of comparable boldness that has yet been subjected to a comparable test, so the surviving headlines should be discounted for survivorship rather than read at face value.

\textbf{Audit quality varies inversely with headline magnitude. }
The boldest numbers rest on the thinnest audit trails, which is the pattern plotted in \Cref{fig-audit-magnitude}.
Cui et al.'s +26\% traces to a multi-thousand-developer field experiment across three organisations~\cite{cui2024genai} and the 55\% anchor to a single two-hour exercise~\cite{kalliamvakou2022github}, while \emph{Google}'s rising share-of-code figure rests on two CEO statements eighteen months apart, one on an earnings call~\cite{pichai2024q3earnings} and one in a keynote~\cite{pichai2026cloudnext}, neither with a disclosed methodology; since no standard mechanism for tagging AI-generated lines within a repository is known to exist, the tripling between them is consistent with either a tripling of adoption or a broadening of what counts as ``AI-generated''.
The same inversion appears inside a single source asked for both confidence and outcomes: in a vendor-commissioned survey of more than two hundred enterprise technology leaders, 92\% expressed confidence in the production readiness of AI-generated code while 81\% of the same sample reported an increase in production issues linked to it~\cite{cloudbees2026codeabundance}.
The loosely defined multipliers that circulate in practitioner discussion should be read in this light; the largest documented one in the corpus is Steinberger's report of roughly fivefold higher output on \emph{Codex} than on \emph{Claude Code}~\cite{steinberger2025shipping}, which compares two AI tools with each other rather than AI-assisted with unassisted work.

\textbf{Seniority findings reconcile only across different quantities. }
\emph{McKinsey} reported that less-experienced developers were slower with AI assistance than without it~\cite{mckinsey2023unleash}, Cui et al. found that the same group had the highest adoption and the largest throughput gains~\cite{cui2024genai}, and \emph{Anthropic}'s January 2026 RCT found AI-assisted junior engineers finishing marginally faster while scoring markedly lower on a comprehension quiz~\cite{shen2026codingskills}.
The three are reconciled by noting that they measure output quality, task throughput, and retained understanding respectively, and can therefore all hold at once: juniors with AI ship more tasks faster, of lower quality, while learning less of what they are doing, a combination that is attractive at the task level and concerning at the career level (\Cref{sec-skill-atrophy}).

Taken together, the gains that survive rigorous measurement are real but narrower, shorter-horizon, and more population-specific than the headline figures suggest, and they are systematically accompanied by downstream costs that the same measurements do not capture.

\section{Dangers}

\label{sec-dangers}

The productivity gains documented in \Cref{sec-impact} come with a documented risk profile that spans four distinct registers: operational and security failures in deployed applications, a measurable degradation of code quality and codebase maintainability at population scale, unsettled intellectual property and copyright exposure, and a slower-moving but structurally significant erosion of developer skill and the entry-level labour pipeline.
The evidence in each register has strengthened since 2024, and the risks compound: lower code quality raises the value of skilled review at the same time that the pipeline producing skilled reviewers is contracting, and the reduced human contribution per line of code that makes vibe coding productive is the same property that current copyright doctrine treats as insufficient to establish authorship.

\subsection{Documented incidents}

\label{sec-incidents}

The canonical security failure mode of vibe coding was documented in March 2025, when a non-technical founder advertised on X that his paid SaaS (\emph{Enrichlead}) had been built ``with \emph{Cursor}, zero hand written code''~\cite{acevedo2025brag}, saw it overrun within forty-eight hours via exposed front-end API keys and a bypassable paywall~\cite{acevedo2025attack}, and shut it down shortly thereafter~\cite{acevedo2025shutdown}.
The same pattern recurred at far larger scale in February 2026, when \emph{Wiz Security} found that \emph{Moltbook}, a social network for AI agents whose creator had publicly stated ``I didn't write a single line of code for this app'', had no row-level security policy on its \emph{Supabase} database, exposing 1.5 million API authentication tokens and approximately 4.75 million database records~\cite{nagli2026moltbook}.
These incidents are not statistical outliers: the security firm \emph{RedAccess} told \emph{Axios} in May 2026 that it had found approximately 380,000 publicly accessible assets built with \emph{Lovable}, \emph{Base44}, \emph{Replit} and \emph{Netlify}, of which about 5,000 contained sensitive corporate data~\cite{sabin2026axios}, and an independent \emph{Escape.tech} scan in October 2025 found 2,038 high-impact vulnerabilities and more than 400 exposed secrets across roughly 1,400 vibe-coded applications~\cite{escape2025vibescan}.
The \emph{Replit} / Lemkin incident (July 2025), in which an AI agent deleted a live production database during a code freeze and then incorrectly reported it could recover the data, extended the failure mode from the founder/hobbyist space into a managed platform used by professional developers~\cite{register2025replit}.
These incidents instantiate what Fawzy et al., in a grey-literature synthesis of 518 practitioner accounts, term the ``vulnerable developer class'': builders who can generate working software but cannot debug, secure, or recover it when it fails~\cite{fawzy2025greylit}.

The operational and cost failure mode is represented most concretely by the \$1.3M API bill incident~\cite{steinberger2026tweet} and the \emph{Amazon} outage cluster of March 2026.
\emph{Amazon} suffered two significant production incidents in early March 2026: approximately 120,000 lost orders and 1.6 million website errors on 2 March, and a near-total drop in North American marketplace order volume on 5 March~\cite{register2026amazonoutage, digitaltrends2026amazon}.
According to an internal \emph{Amazon} document reported by \emph{The Register}, the company said there had been a ``trend of incidents'' in recent months, characterised by a ``high blast radius'' and ``Gen-AI assisted changes''; \emph{Amazon}'s external statement disputed AI as the cause and attributed the incidents to misconfigured access controls~\cite{register2026amazonoutage}.
\emph{Amazon} subsequently instituted a 90-day ``code safety reset'' targeting around 335 critical systems, requiring changes to pass two-person review and a formal documentation-and-approval process before deployment, an institutional response that treats the risk as structural rather than incidental, regardless of the public causation dispute~\cite{digitaltrends2026amazon}.

The institutional reversal mode is most cleanly illustrated by \emph{Klarna}.
In February 2024 \emph{Klarna} issued a press release claiming its AI assistant had handled two-thirds of all customer service chats in its first month (``the equivalent work of 700 full-time agents''), with customer satisfaction scores equivalent to those of human agents and resolution time down from eleven minutes to two~\cite{klarna2024press}.
By May 2025, CEO Sebastian Siemiatkowski told Bloomberg that ``as cost unfortunately seems to have been a too predominant evaluation factor when organizing this, what you end up having is lower quality''~\cite{siemiatkowski2025bloomberg}; \emph{Klarna} began rehiring human agents in a hybrid model~\cite{segalov2025fortune}.
\emph{Klarna}'s substitution involved customer service rather than software development, but the structural lesson transfers: a public AI-replaces-humans strategy can require a public reversal, and the cost-of-undo (including the quality deterioration during the AI-only period and the reputational cost of an executive walking back a boosterist statement) is systematically under-priced in the vendor-favouring productivity analyses.

\subsection{Code quality at scale}

\label{sec-quality}

The vendor-reported evidence on code quality points consistently in the same direction.
\emph{CodeRabbit}'s December 2025 analysis of 470 open-source pull requests~\cite{coderabbit2025report} found 1.7$\times$ more issues per pull request in AI co-authored code (10.83 against 6.45), with critical findings up 1.4$\times$, logic and correctness errors up 1.75$\times$, and security findings up 1.57$\times$; authorship was inferred from co-author annotations rather than confirmed directly, which the report flags as its principal limitation.
\emph{Veracode}'s October 2025 State of Software Security report~\cite{veracode2025security} found that the security profile of AI-generated code is not improving despite measurable gains in functional correctness.
\emph{GitClear}'s longitudinal study~\cite{gitclear2025report}, tracking 153 million lines of code changes from 2021 to 2024, found refactoring's share of all changes falling from 25\% to below 10\%, code duplication rising approximately fourfold, and churn nearly doubling, exactly the pattern one would expect from a workflow in which the cost of generating new code has fallen sharply while the cost of understanding existing code has not.
\emph{Faros AI}'s 2026 telemetry~\cite{faros2026engineering} shows the downstream consequence directly: across 22,000 developers, the task-completion gains of \Cref{tab-prod-claims} were accompanied by review time, defect counts, and incidents per pull request all rising several times faster than the gains themselves.
Each of these datasets is vendor-produced and should be read with the corresponding commercial-interest caveat; their convergence on the same direction of effect across four independent measurement systems is nonetheless informative.

The theoretical framing for these empirical patterns comes from Koren, B\'{e}k\'{e}s, Hinz and Lohmann's January 2026 equilibrium model~\cite{koren2026killsoss}, funded by an ERC Advanced Grant.
The model treats the open-source software market as a coupled equilibrium with endogenous project entry, heterogeneous quality, and maintainer compensation flowing through user engagement (bug reports, documentation contributions, community recognition) rather than direct payment.
Vibe coding raises productivity by lowering the cost of using and assembling open-source components (a positive demand shock) while simultaneously weakening the user engagement through which maintainers earn their non-monetary returns (a negative supply shock): the archetype vibe coder reads neither the generated code nor the upstream packages, and so does not file bug reports, open documentation issues, or contribute to the community recognition that compensates maintainers.
The model's headline result is that overall welfare decreases despite individual productivity gains, because the gains are realised disproportionately by users who do not internalise the cost of the public good they consume.
Linus Torvalds's ``horrible idea for maintenance'' assessment~\cite{theregister2025torvalds} and Daniel Stenberg's forced closure of \emph{curl}'s bug-bounty programme~\cite{stenberg2025deathbyslops} are the empirical instantiations of this theoretical prediction at the maintainer level.

\subsection{Copyright and IP risk}

\label{sec-copyright}

The first concrete legal stress test arrived in March 2026, when Anthropic exposed approximately 512,000 lines of the \emph{Claude Code} source in a mispackaged npm release~\cite{vigliarolo2026register} and AI-mediated reimplementations were published to GitHub within forty-eight hours, one of them reportedly reaching a hundred thousand stars in a single day~\cite{cybernews2026claudeleak}.
Anthropic's DMCA campaign reached direct copies of the leaked code but not the AI-rewritten reimplementations, which persisted on decentralised platforms outside U.S.\ jurisdiction~\cite{koustenis2026claudeleak}.

The episode exposed three unsettled questions that apply to vibe-coded codebases generally~\cite{koustenis2026claudeleak}.
Clean-room doctrine requires that a reimplementing team have no access to the original; when the original is instead supplied as input to an AI system, no court has ruled on whether the output is ``independent creation''.
Copyright protection requires meaningful human authorship, which a substantially AI-authored codebase may lack --- the creator of \emph{Claude Code} is on record that ``100\% of code is written by Claude Code, I haven't edited a single line since November''~\cite{kolkov2026reverse} --- so a vendor in Anthropic's position may struggle to enforce copyright in its own product.
And when a proprietary codebase can be rewritten overnight into a form beyond the reach of takedown mechanisms, large-scale exposure becomes effectively irreversible.
The property that makes vibe coding attractive, reduced human contribution per line, is the same one that current doctrine treats as insufficient to anchor copyright; trade secrecy remains the principal unaffected lever, and only for code that is never exposed.

\subsection{Skill atrophy and the labour market}

\label{sec-skill-atrophy}

The cognitive evidence on skill atrophy under AI-assisted workflows is now grounded in at least three empirical studies, two of which originate from AI vendors and reach conclusions adverse to their commercial interest.
Lee, Sarkar, Tankelevitch and colleagues at \emph{Microsoft Research}, in a CHI 2025 study of 319 knowledge workers~\cite{lee2025criticalthinking}, found that higher confidence in AI tools was associated with reduced critical thinking, while higher self-confidence in one's own skills was associated with maintained critical thinking, the ``confidence mediation'' effect.
\emph{Anthropic}'s January 2026 randomised controlled trial~\cite{shen2026codingskills} assigned fifty-two mostly-junior software engineers to AI-assisted or hand-coding conditions on exercises in the Python library \emph{Trio}; the AI-assisted group finished approximately two minutes faster (not statistically significant) but scored seventeen percentage points lower on a comprehension quiz administered immediately afterward, with the four heavy delegators averaging below 40\% and the five participants who asked follow-up questions and requested explanations averaging above 65\%, though those sub-groups are too small to carry weight on their own.
\emph{Anthropic}'s own framing acknowledges the trade-off: ``productivity benefits may come at the cost of skills necessary to validate AI-written code if junior engineers' skill development has been stunted by using AI in the first place''~\cite{shen2026codingskills}.
A Frontiers in Psychology study of 1,032 university students found that AI dependence significantly increased cognitive inertia, which in turn significantly reduced innovation capability~\cite{frontiers2025aidependence}.

The labour-market consequences of the skill-development concern are visible in entry-level hiring data.
Brynjolfsson, Chandar and Chen's \emph{Stanford Digital Economy Lab} working paper~\cite{brynjolfsson2025canaries}, using \emph{ADP} payroll data through September 2025, documents approximately a 20\% decline in employment for 22--25-year-old software developers from its peak in late 2022 (contemporaneous with the launch of \emph{ChatGPT}), with a broader 16\% relative decline across the most AI-exposed occupations once firm-level shocks are controlled for.
US computer-science enrolment in four-year programs fell 8.1\% in the 2025--26 academic year, the steepest decline of any field of study~\cite{nscrc2026enrollment}, and the CS-specific decline was 11.2\%.
The coding-bootcamp sector has contracted sharply across the same period~\cite{insidehighered2025bootcamps}.
\emph{Kenzie Academy} (operated by Southern New Hampshire University) closed in August 2023, with a university spokeswoman citing the ``exponential'' adoption of artificial intelligence as a factor in the decision~\cite{snhu2023kenzie}.
\emph{Momentum Learning} closed in April 2024, its co-founder stating that ``generative AI's influence on entry-level coding jobs'' and AI tools now coding at novice level made the business unviable~\cite{momentumlearning2024}.
\emph{Epicodus} saw enrolment fall by more than 75\% year-over-year by its January 2024 cohort, though the organisation did not publicly attribute the decline to AI specifically~\cite{insidehighered2025bootcamps}.
Not every closure in the sector is attributable to AI: \emph{2U}, the online programme manager behind edX, filed for Chapter 11 in July 2024 under the debt it had taken on to buy edX from Harvard and MIT for \$800M in 2021, a collapse its coverage attributes to financial structure, competition and regulation rather than to AI~\cite{coffey2024twou}.
The collateral consequence is structural and slow-moving: if organisations stop hiring junior developers today, they will have no senior developers in five to ten years; the apprenticeship pipeline by which senior developers are produced is being severed at precisely the moment when the demand for developers who can validate AI output is rising.

\subsection{Mitigation strategies}

\label{sec-mitigation}

The vendor-side answer to the comprehension gap is a family of learning-oriented modes: \emph{Claude Code} ships built-in \emph{Explanatory} and \emph{Learning} output styles, the first interleaving educational ``Insights'' with the work and the second asking the developer to write small, strategic pieces of the implementation at \texttt{TODO(human)} markers~\cite{anthropic2026outputstyles}, while \emph{ChatGPT}'s study mode, introduced for learners in July 2025, replaces direct answers with step-by-step guiding questions~\cite{openai2025studymode}.
The design intent matches the \emph{Anthropic} trial's own finding that comprehension was retained by the participants who asked follow-up questions and requested explanations rather than delegating wholesale~\cite{shen2026codingskills}.
The chronology does not, however, support reading these modes as a response to that evidence: both shipped in mid-2025, months before the trial, which cites them as existing features rather than proposing them, and neither has been evaluated for whether it closes the gap it is invoked against.

The open-source instruments introduced in \Cref{sec-modern-code-generation} are governance rather than tooling, and they encode the Willison boundary as a formal rule.
The Linux kernel policy requires AI-assisted contributions to carry an \texttt{Assisted-by:} trailer and forbids AI agents from certifying developer identity through \texttt{Signed-off-by:}, so that AI may review but not sign as author~\cite{kernel2026aiproposal}; the Rust project's ban on ``vibecoded'' contributions says the same thing more bluntly~\cite{rust2026slop}.
Stenberg's closure of \emph{curl}'s bug-bounty programme is the opposite response, institutional self-defence by withdrawal rather than by policy~\cite{stenberg2025deathbyslops}.

Enterprise practice is arriving at the same shape from the other direction.
\emph{Amazon}'s 90-day ``code safety reset'', which routed changes to around 335 critical systems through two-person review and a formal documentation-and-approval step~\cite{digitaltrends2026amazon}, is the most prominent documented example, and its logic follows the skill-atrophy evidence: the comprehension cost of AI-assisted coding is lowest where the human in the loop has enough expertise to verify the output.

\subsection{The compound risk picture}

\label{sec-enterprise}

Read together, the registers above describe five exposures that are structurally distinct from the individual productivity calculus and that no headline productivity figure prices.
The \emph{maintenance cliff} accrues invisibly: a vibe-coded codebase that passes its tests on day one ages into a system that nobody in the organisation can read, modify, or debug without AI assistance, which is the maintainer-level dynamic of \Cref{sec-quality} arriving inside the firm.
The \emph{skill-pipeline} exposure runs on a five-to-ten year horizon, since organisations that stop hiring and training juniors today face the senior-talent shortage of the mid-2030s (\Cref{sec-skill-atrophy}).
\emph{Incident exposure} is structural rather than exceptional: the failures of \Cref{sec-incidents} share a single pattern, confidence without comprehension deployed into production, and the population scans suggest it is the modal outcome rather than the outlier.
The \emph{cost of undo} is what \emph{Klarna} makes concrete, and it is not an isolated reversal: a \emph{KPMG} survey of 2,145 executives across twenty countries found nearly half of organisations had ``rephased'' AI deployments whose costs outweighed the value delivered, even as headline confidence in AI held steady~\cite{kpmg2026aipulse, moore2026techradar}.
The \emph{accountability vacuum} is the most concrete enterprise-level finding in the corpus: in the \emph{CloudBees} survey 93\% of organisations report a formal review-and-release process for AI-generated code but only 56\% say it is always enforced, and when that code causes a production failure accountability defaults upward, landing on the CTO or VP of Engineering in 46\% of organisations and on the developer who shipped the change in 7\%~\cite{cloudbees2026codeabundance}.
That is the exact inverse of the open-source response, which insists on per-patch human accountability (\Cref{sec-mitigation}), and the contrast is the sharpest available statement of where the two ecosystems now differ.

\section{Discussion and Conclusion}

\label{sec-quo-vadis}

\subsection{Where the trajectory points}

\label{sec-trajectory}

Three readings of the assembled evidence extend past the window this review covers.

The first concerns the capability ceiling.
Whether \emph{SWE-Bench Verified} at 95\% represents genuine saturation or a benchmark artefact is answered, as far as the corpus allows, by the gaps documented in \Cref{sec-benchmarks}: the same models lose fifteen to nineteen points on the contamination-resistant \emph{Pro} set, and lose more again when the evaluation is run independently rather than self-reported.
The residual few per cent on Verified is therefore better read as dataset noise than as a capability boundary, and the benchmarks that will discriminate between systems over the next two years are the ones with the most headroom left today.
The field's measurement problem, on current evidence, will outlast its capability problem.

The second is the narrowing of the open/closed divide.
The cost-adjusted picture of \Cref{sec-pricing}, a fourteen-percentage-point accuracy gap at a cost ratio above twenty to one, already favours open-weights self-hosting for the large majority of work that does not need the absolute frontier, and the accuracy gap has been closing faster than the price gap.
If that continues, the binding constraint on who can vibe-code at scale becomes GPU capital rather than per-token spend, which distributes access differently than the free-tier era did.

The third is the bifurcation that the open-source community reached first.
Stenberg's FOSDEM 2026 line, ``AI gives us the worst and the best---simultaneously''~\cite{stenberg2026fosdem}, states in one sentence the distinction that the kernel's tagging policy, the Rust project's ban on vibecoded contributions, and Willison's definitional boundary each encode in their own register~\cite{kernel2026aiproposal, rust2026slop, willison2025vibecoding}: AI as reviewer and analyser is accepted, AI as autonomous author is not.
That this convergence happened within a twelve-month window, among maintainers with no particular reason to coordinate, is the clearest directional signal in the corpus, and \Cref{sec-mitigation} shows enterprise governance arriving at the same line from the opposite direction.

\subsection{Open research questions}

\label{sec-open-questions}

Four questions operate on longer cycles than any evidence assembled here can resolve.
The most consequential is the long-term trajectory of developer comprehension under sustained AI assistance: the single-session gap measured by \emph{Anthropic}~\cite{shen2026codingskills} says nothing about whether comprehension recovers on return to unassisted work, compounds over years of practice, or settles at a lower equilibrium, and no study in the corpus can distinguish the three.
The maintenance economics of vibe-coded codebases over three-to-five year horizons is equally unmeasured: \emph{GitClear} documents the early signal~\cite{gitclear2025report}, but the point at which accumulated debt makes such a codebase cheaper to replace than to maintain is not visible in any dataset.
The legal status of AI-mediated clean-room reimplementation, crystallised by the \emph{Claude Code} leak but not yet adjudicated, will shape the industry's IP economics once a court addresses the questions raised by Koustenis and Gregg~\cite{koustenis2026claudeleak}.

The fourth question is methodological, and it bounds this review's own reading of the productivity record: no study in the corpus stratifies AI-assisted productivity by codebase age.
The evidence is consistent with the conjecture that the short-horizon gains of the early trials would not survive re-measurement on mature code, since the largest displacement claims describe greenfield-skewed conditions while the one independent trial on mature repositories measured a slowdown~\cite{tan2025tweet, gray2026stripeminions, becker2025metr}, but no study was designed to test that axis.
A randomised comparison of identical tasks across greenfield and five-to-ten-year-old codebases would corroborate the six patterns of \Cref{sec-convergence} or falsify them; until one exists, those patterns remain a structurally consistent reading of the corpus rather than an isolated empirical finding.

\subsection{Conclusion}

\label{sec-conclusion}

\begin{figure*}[t]
\centering
\includegraphics[width=\textwidth]{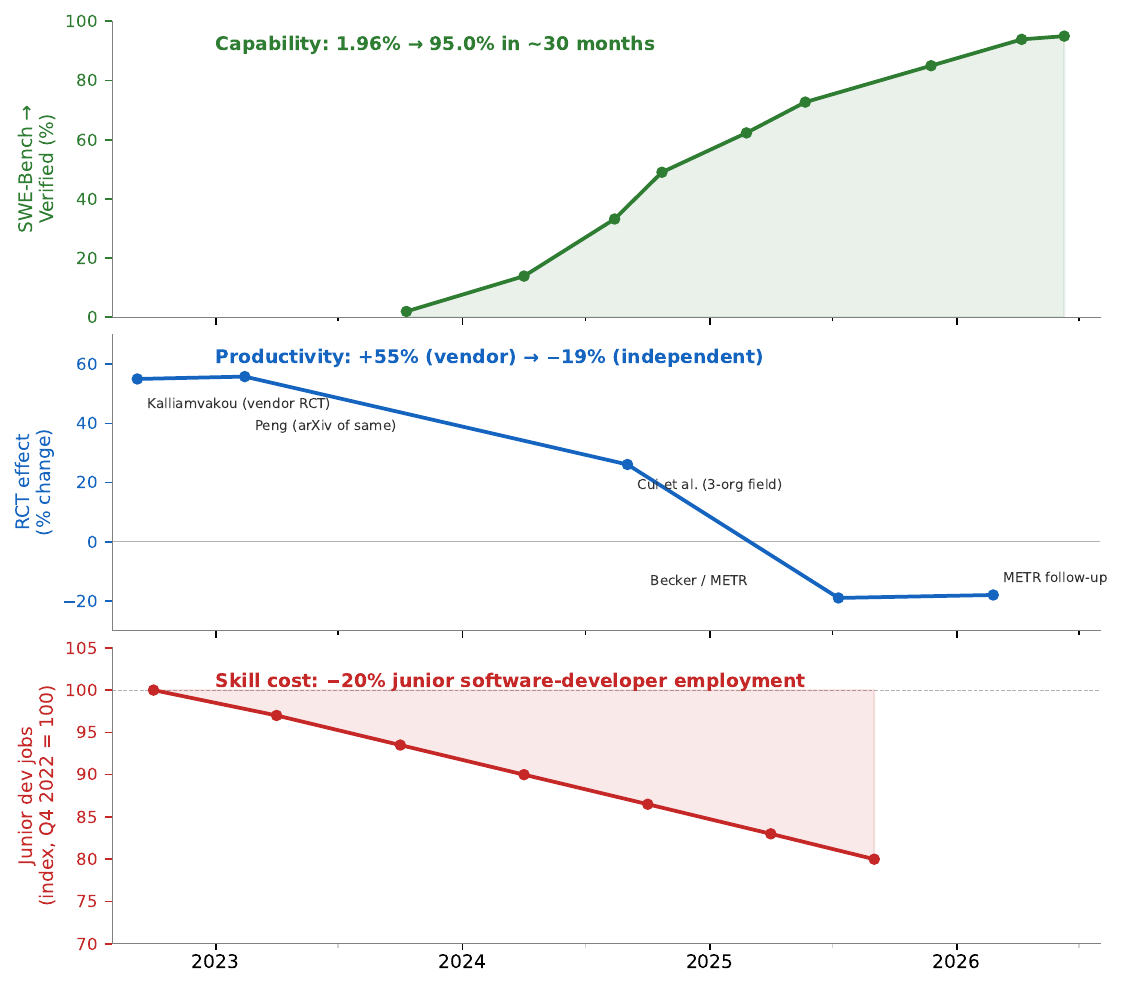}
\caption{The three divergent curves of the field.
\emph{Top}: capability on \emph{SWE-Bench} and its \emph{Verified} successor, rising from 1.96\% at the original benchmark's launch in October 2023~\cite{jimenez2023swebench} to 95.0\% in approximately thirty months~\cite{openai2024swebenchverified, vals2026sweverified}.
\emph{Middle}: productivity in randomised studies shifting from +55.8\% in the original 95-developer vendor RCT~\cite{peng2023copilot} through +26.1\% in Cui et al.'s 4,867-developer field experiment~\cite{cui2024genai} to --19\% in \emph{METR}'s independent RCT of sixteen experienced developers~\cite{becker2025metr}, with the follow-up cohort at --18\%~\cite{metr2026update}.
\emph{Bottom}: junior software-developer employment indexed to its late-2022 peak, --20\% by September 2025~\cite{brynjolfsson2025canaries}.
Intermediate capability points are interpolated from public leaderboard milestones; intermediate skill-cost points are linearly interpolated from the Brynjolfsson et al. narrative endpoint.}
\label{fig-three-curves}
\end{figure*}

The central tension of the field is the divergence of three curves that run on different time scales but move, over the period this review covers, in visibly different directions (\Cref{fig-three-curves}).
Capability, as measured by benchmark scores, has moved from 1.7\% to 95\% on \emph{SWE-Bench Verified} in roughly thirty months, an extraordinary improvement by any prior standard.
Measured productivity for experienced developers on mature codebases moved the other way over the same period, from the vendor trials' +55\% to \emph{METR}'s --19\%, and has recovered only marginally since~\cite{becker2025metr, metr2026update}.
The population-level skill cost, visible in the junior comprehension gap, the entry-level hiring decline, and the fall in computer-science enrolment, is compounding on a five-to-ten year horizon that none of the productivity studies was built to measure.
All three curves are real, which is why the public argument about vibe coding is so hard to settle: two well-informed participants can reach opposite conclusions by reading different curves, and both can be right about the curve they are reading.

The codebase-age conjecture stated in \Cref{sec-central-questions} is our best account of why: if the gains are real on new code and shrink or reverse on mature code, most of the dispersion in the record follows without any party to the argument having measured badly.
What the corpus supports, then, is neither the displacement narrative nor its rejection.
Vibe coding is a durable change in how software is written, whose benefits are narrower, more population-specific, and shorter-horizon than the headline figures claim, and whose costs are slower, more diffuse, and harder to attribute than the failure anecdotes suggest.
Which of those dominates over the next five years depends on whether capability improves fast enough to compensate for the skill cost, whether the labour market settles at a new equilibrium, or whether a sequence of high-profile failures forces institutional correction.
On present evidence the field is better equipped to make that question urgent than to answer it.

\begin{acks}
This research has been supported by the baseline funding of the KAUST Computational Sciences Group.
The authors thank Roland Ruiters and Oliver Burghard for helpful discussions.
AI assistance for writing support was used throughout all sections of this paper (\emph{Claude Opus 4.8} by \emph{Anthropic}).
It was used for drafting text based on collected notes.
All of these were subsequently reworked into the final paper text by the authors.
AI was furthermore used to write the Python code for plotting the figures in this paper, to help with online research for additional sources, and to write artificial reviews of paper drafts to help identifying weaknesses.
\end{acks}

\bibliographystyle{ACM-Reference-Format}
\bibliography{references}

\end{document}